\documentclass[11pt]{article}
\usepackage{amssymb,amsmath,amsfonts}
\usepackage{graphicx}
\usepackage{graphics}
\usepackage{eepic,epsfig}

\begin{document}

\title{Plane symmetric gravitational fields in (D+1)-dimensional \\
General Relativity}
\author{R. M. Avagyan$^{1,2}$, \thinspace\ T. A. Petrosyan$^{1}$,\thinspace\
A. A. Saharian$^{1,2}$\thanks{%
Corresponding author, e-mail: saharian@ysu.am}, G. H. Harutyunyan$^{1}$ \\
\\
\textit{$^1$Institute of Physics, Yerevan State University,}\\
\textit{1 Alex Manoogian Street, 0025 Yerevan, Armenia} \vspace{0.3cm}\\
\textit{$^2$Institute of Applied Problems of Physics NAS RA,}\\
\textit{25 Hrachya Nersissyan Street, 0014 Yerevan, Armenia}}
\maketitle

\begin{abstract}
We consider plane symmetric gravitational fields within the framework of
General Relativity in (D+1)-dimensional spacetime. Two classes of vacuum
solutions correspond to higher-dimensional generalizations of the Rindler
and Taub spacetimes. The general solutions are presented for a positive and
negative cosmological constant as the only source of the gravity. Matching
conditions on a planar boundary between two regions with distinct plane
symmetric metric tensors are discussed. An example is considered with
Rindler and Taub geometries in neighboring half-spaces. As another example,
we discuss a finite thickness cosmological constant slab embedded into the
Minkowski, Rindler and Taub spacetimes. The corresponding surface
energy-momentum tensor is found required for matching the exterior and
interior geometries.
\end{abstract}

\bigskip

\section{Introduction}

Exact solutions of Einstein's equations for the gravitational field are
available only in geometries with relatively high symmetry (for reviews see 
\cite{Step03,Grif09}). In particular, they include spherical, axial and
planar symmetric configurations. Another classes of solutions with maximally
symmetric subspaces are used in cosmology. Despite their apparent
simplicity, plane symmetric solutions remain an active subject of research.
The investigations are motivated by interesting geometrical properties of
those solutions and by their applications in different areas of
gravitational physics. The latter include the domain wall type topological
defects in field theories \cite{Vile94} and branes in string theory and in
braneworld models with extra dimensions \cite{John09,West12,Maar10}.

The static plane-symmetric vacuum solutions of Einstein's equations were
already known in the early days of the development of the General Relativity 
\cite{Levi18} and were rediscovered later in \cite{Taub51}. Two classes of
solutions are present. The first one corresponds to the Rindler spacetime
and describes a flat geometry. It approximates the gravitational field near
the black hole horizon and is among the most popular geometries in quantum
field theory on backgrounds with horizons (see, for example, \cite{Birr82}).
The second class of single parameter solutions corresponds to the Taub
geometry. The characteristic feature of the latter is the presence of a
curvature singularity on a plane with a fixed value of the coordinate along
which the geometry is inhomogeneous. The test particle is repelled by the
singularity. The nature of the singularity, the other properties of the Taub
solution and its generalizations in the presence of the matter sources have
been widely discussed in the literature (see, e.g., \cite{Taub56}-\cite%
{Hali23} and references therein). In the present paper we discuss several
aspects of plane-symmetric static solutions in $(D+1)$-dimensional General
Relativity. Higher dimensional gravitational configurations with planar
symmetry appear in a number of models including braneworld scenarios,
anti-de Sitter/Conformal field theory (AdS/CFT) correspondence and
fundamental branes in string theories and supergravity.

The organization of the paper is as follows. In the next section, the
background geometry, gravitational field equations and the matching
conditions in problems with different metric tensors in separate regions are
presented. In Section \ref{sec:Vacsol} two classes of vacuum solutions
corresponding to the Rindler and Taub spacetimes are considered. The
solutions with a cosmological constant (CC) as the only source in the
gravitational field equations are discussed in Section \ref{sec:CCsol}. In
Section \ref{sec:CCslab} we consider a slab with CC interior and with
different exterior geometries. The corresponding surface energy-momentum
tensors required by the matching conditions on the slab boundaries are given.

\section{Background geometry, the field equations and matching conditions}

\label{sec:Geom}

We consider a plane symmetric $(D+1)$-dimensional spacetime with the line
element%
\begin{equation}
ds^{2}=e^{2u_{0}}dt^{2}-e^{2u_{1}}dx^{2}-e^{2u_{2}}\sum_{i=2}^{D}\left(
dx^{i}\right) ^{2},  \label{ds2s}
\end{equation}%
where $x^{1}=x$, $u_{l}=u_{l}(x)$, $l=0,1,2$. The nonzero components of the
Ricci tensor are given by the expressions%
\begin{eqnarray}
R_{0}^{0} &=&e^{-2u_{1}}\left[ u_{0}^{\prime \prime }+u_{0}^{\prime
2}-u_{0}^{\prime }u_{1}^{\prime }+(D-1)u_{0}^{\prime }u_{2}^{\prime }\right]
,  \notag \\
R_{1}^{1} &=&e^{-2u_{1}}\left[ u_{0}^{\prime \prime }+u_{0}^{\prime
2}-u_{1}^{\prime }u_{0}^{\prime }-(D-1)\left( u_{2}^{\prime \prime
}+u_{2}^{\prime 2}-u_{1}^{\prime }u_{2}^{\prime }\right) \right] ,  \notag \\
R_{2}^{2} &=&e^{-2u_{1}}\left[ u_{2}^{\prime \prime }+u_{0}^{\prime
}u_{2}^{\prime }-u_{1}^{\prime }u_{2}^{\prime }+(D-1)u_{2}^{\prime 2}\right]
,  \label{R22}
\end{eqnarray}%
and (no summation over $i$) $R_{i}^{i}=R_{2}^{2}$ for $i=3,\ldots ,D$. Here,
the prime stands for the derivative with respect to the coordinate $x$. The
Ricci scalar is expressed as 
\begin{equation}
R=2e^{-2u_{1}}\left[ u_{0}^{\prime \prime }+u_{0}^{\prime 2}-u_{0}^{\prime
}u_{1}^{\prime }+(D-1)\left( u_{2}^{\prime \prime }+u_{0}^{\prime
}u_{2}^{\prime }-u_{1}^{\prime }u_{2}^{\prime }+\frac{D}{2}u_{2}^{\prime
2}\right) \right] .  \label{Ric}
\end{equation}%
For discussion of geodesic motion one needs also to have the Christoffel
symbols. The expressions for the corresponding nonzero components read (no
summation over $i=2,3,\ldots ,D$)%
\begin{eqnarray}
\Gamma _{01}^{0} &=&\Gamma _{10}^{0}=u_{0}^{\prime },\;\Gamma
_{00}^{1}=e^{2\left( u_{0}-u_{1}\right) }u_{0}^{\prime },\;\Gamma
_{11}^{1}=u_{1}^{\prime },  \notag \\
\Gamma _{ii}^{1} &=&-e^{2\left( u_{2}-u_{1}\right) }u_{2}^{\prime },\;\Gamma
_{1i}^{i}=\Gamma _{i1}^{i}=u_{2}^{\prime }.  \label{Chris}
\end{eqnarray}%
The $i$th component of the acceleration for a test particle is given by $%
a^{i}=-\Gamma _{kl}^{i}w^{k}w^{l}$ with $w^{i}=dx^{i}/ds$ being the $(D+1)$%
-velocity. For a test particle at rest one has $w^{i}=\delta
_{0}^{i}e^{-u_{0}}$ and the acceleration is directed along the $x$-axis with 
$a^{i}=d^{2}x^{i}/ds^{2}=-\delta _{1}^{i}e^{-2u_{1}}u_{0}^{\prime }$.

For the gravitational field equations $R_{i}^{k}-\delta _{i}^{k}R/2=8\pi
GT_{i}^{k}$, with $T_{i}^{k}$ being the metric energy-momentum tensor, we get%
\begin{eqnarray}
-\frac{8\pi GT_{0}^{0}e^{2u_{1}}}{D-1} &=&u_{2}^{\prime \prime }+\frac{D}{2}%
u_{2}^{\prime 2}-u_{1}^{\prime }u_{2}^{\prime },  \notag \\
-\frac{8\pi GT_{1}^{1}e^{2u_{1}}}{D-1} &=&u_{2}^{\prime }\left(
u_{0}^{\prime }+\frac{D-2}{2}u_{2}^{\prime }\right) ,  \notag \\
-8\pi GT_{2}^{2}e^{2u_{1}} &=&u_{0}^{\prime \prime }+u_{0}^{\prime
2}-u_{0}^{\prime }u_{1}^{\prime }+(D-2)\left( u_{2}^{\prime \prime
}+u_{0}^{\prime }u_{2}^{\prime }-u_{1}^{\prime }u_{2}^{\prime }+\frac{D-1}{2}%
u_{2}^{\prime 2}\right) .  \label{T22}
\end{eqnarray}%
In accordance with the problem symmetry one has (no summation over $i$) $%
T_{i}^{i}=T_{2}^{2}$ for $i=3,\ldots ,D$. Note that the quantity $-T_{i}^{i}$%
, $i=1,2,\ldots ,D$, presents the effective pressure along the $i$th spatial
dimension. From the covariant conservation equation $\nabla _{k}T_{i}^{k}=0$
one finds%
\begin{equation}
T_{1}^{1\prime }+\left( T_{1}^{1}-T_{0}^{0}\right) u_{0}^{\prime
}+(D-1)\left( T_{1}^{1}-T_{2}^{2}\right) u_{2}^{\prime }=0.  \label{Cons}
\end{equation}%
This equation does not contain the function $u_{1}(x)$. For a source with
barotropic equation of state, $T_{i}^{i}=-w_{i}T_{0}^{0}$ with constants $%
w_{i}$, we get 
\begin{equation}
T_{1}^{1}=\mathrm{const}\cdot \exp \left[ -\left( \frac{1}{w_{1}}+1\right)
u_{0}+(D-1)\left( \frac{w_{2}}{w_{1}}-1\right) u_{2}\right] .  \label{TiiBE}
\end{equation}%
In this case the components of the energy-momentum tensor, as functions of
the coordinate $x$, do not change the sign.\qquad

The function $u_{1}(x)$ in (\ref{ds2s}) can be fixed by the choice of the
coordinate $x$. The field equations are essentially simplified taking 
\begin{equation}
u_{1}(x)=0.  \label{u10}
\end{equation}%
This gives%
\begin{eqnarray}
-\frac{8\pi GT_{0}^{0}}{D-1} &=&u_{2}^{\prime \prime }+\frac{D}{2}%
u_{2}^{\prime 2},  \notag \\
-\frac{8\pi GT_{1}^{1}}{D-1} &=&u_{2}^{\prime }\left( u_{0}^{\prime }+\frac{%
D-2}{2}u_{2}^{\prime }\right) ,  \notag \\
-8\pi GT_{2}^{2} &=&u_{0}^{\prime \prime }+u_{0}^{\prime 2}+(D-2)\left(
u_{2}^{\prime \prime }+u_{0}^{\prime }u_{2}^{\prime }+\frac{D-1}{2}%
u_{2}^{\prime 2}\right) .  \label{T22b}
\end{eqnarray}%
From these equations the following relations can be obtained:%
\begin{eqnarray}
\left[ u_{0}^{\prime }e^{u_{0}+(D-1)u_{2}}\right] ^{\prime } &=&\frac{8\pi G%
}{D-1}\left[ (D-2)T_{0}^{0}-T_{1}^{1}-(D-1)T_{2}^{2}\right]
e^{u_{0}+(D-1)u_{2}},  \notag \\
\left[ u_{2}^{\prime }e^{u_{0}+(D-1)u_{2}}\right] ^{\prime } &=&-\frac{8\pi G%
}{D-1}\left( T_{0}^{0}+T_{1}^{1}\right) e^{u_{0}+(D-1)u_{2}}.  \label{Relu01}
\end{eqnarray}%
Note that $e^{u_{0}+(D-1)u_{2}}=\sqrt{|g|}$ with $g$ being the determinant
of the metric tensor $g_{ik}$. Additionally, by combining the equations (\ref%
{Relu01}) we get%
\begin{equation}
\left[ e^{u_{0}+(D-1)u_{2}}\right] ^{\prime \prime }=-\frac{8\pi G}{D-1}%
\left[ T_{0}^{0}+DT_{1}^{1}+(D-1)T_{2}^{2}\right] e^{u_{0}+(D-1)u_{2}}.
\label{Relu01b}
\end{equation}%
The integration of relations (\ref{Relu01}) and (\ref{Relu01b}) give
conditions for the energy-momentum tensor to be compatible with given
solutions for $u_{0}(x)$ and $u_{2}(x)$.

By using the set of equations (\ref{T22b}) we can derive the matching
conditions for the components of the metric tensor in problems where the
geometry is described by two distinct metric tensors in regions separated by
a planar boundary. As a separating boundary we take a hyperplane $x=L$. The
energy-momentum tensor is decomposed into two contributions: 
\begin{equation}
T_{i}^{k}=T_{\mathrm{(v)}i}^{k}+T_{\mathrm{(s)}i}^{k},\;T_{\mathrm{(s)}%
i}^{k}=\tau _{i}^{k}\delta (x-L).  \label{Tikdec}
\end{equation}%
Here, $T_{(v)i}^{k}$ is the volume part and $T_{(s)i}^{k}$ corresponds to
the surface energy-momentum tensor localized on the interface $x=L$.
Generally, the volume part is different in the regions $x<L$ and $x>L$.
Assuming that the metric tensor is continuous at $x=L$, the discontinuities
in its first order derivatives are found by integrating the equations (\ref%
{T22b}) in the region $[L-\varepsilon ,L+\varepsilon ]$, $\varepsilon >0$,
and then taking the limit $\varepsilon \rightarrow 0$. The continuity
conditions for the metric tensor read%
\begin{equation}
u_{0}|_{L-}^{L+}=\lim_{\varepsilon \rightarrow 0}\left[ u_{0}(L+\varepsilon
)-u_{0}(L-\varepsilon )\right] =0,\;u_{2}|_{L-}^{L+}=0.  \label{u02cont}
\end{equation}%
Under these conditions, by taking into account that $\lim_{\varepsilon
\rightarrow 0}\int_{L-\varepsilon }^{L+\varepsilon }dx\,u_{l}^{\prime 2}=0$
and $\lim_{\varepsilon \rightarrow 0}\int_{L-\varepsilon }^{L+\varepsilon
}dx\,u_{0}^{\prime }u_{2}^{\prime }=0$, for the first order derivatives we
get%
\begin{eqnarray}
u_{0}^{\prime }|_{L-}^{L+} &=&8\pi G\left( \frac{D-2}{D-1}\tau _{0}^{0}-\tau
_{2}^{2}\right) ,  \notag \\
u_{2}^{\prime }|_{L-}^{L+} &=&-\frac{8\pi G}{D-1}\tau _{0}^{0},\;\tau
_{1}^{1}=0.  \label{u02der}
\end{eqnarray}%
The discontinuities in the derivatives of the metric tensor are completely
determined by the surface energy-momentum tensor. The corresponding
conditions can also be obtained from the Israel matching conditions in terms
of the extrinsic curvature tensor of the separating boundary.

\section{Vacuum solutions}

\label{sec:Vacsol}

We start with the vacuum solutions of the set of equations (\ref{T22b}). For
them one has $T_{i}^{k}=0$. By having the coordinate $x$ fixed by the
condition (\ref{u10}), we have two possibilities. For the first one $%
u_{2}^{\prime }=0$ and the first and second equations in (\ref{T22b}) are
satisfied identically. From the last equation we get $u_{0}^{\prime \prime
}+u_{0}^{\prime 2}=0$. The solution $u_{0}^{\prime }=0$ corresponds to a
flat spacetime in the Minkowskian coordinates. The solution for $%
u_{0}^{\prime }\neq 0$ is obtained after a simple integration: $%
e^{2u_{0}}=\left( x+C\right) ^{2}$. Taking $C=0$ we get the line element 
\begin{equation}
ds_{\mathrm{R}}^{2}=x^{2}dt_{\mathrm{R}}^{2}-dx^{2}-\sum_{i=2}^{D}\left(
dx^{i}\right) ^{2},  \label{dsR}
\end{equation}%
which corresponds to the Rindler spacetime. Note that in the representation (%
\ref{dsR}) the Rindler time coordinate $t_{\mathrm{R}}$ is dimensionless.
Introducing new coordinates $(T,X)$ in accordance with $T=x\sinh t_{\mathrm{R%
}}$, $X=\mathrm{sgn}(X)\,x\cosh t_{\mathrm{R}}$, the line element (\ref{dsR}%
) takes the Minkowskian form. The coordinates $(t_{\mathrm{R}%
},x,x^{2},\ldots ,x^{D})$ cover the Rindler wedges $|X|>|T|$ of the
Minkowski spacetime. The worldline with fixed $(x,x^{2},\ldots ,x^{D})$
describes a uniformly accelerated observer having the proper acceleration $%
1/x$. The hypersurface $x=0$ corresponds to the Rindler horizon.

For the second class of the vacuum solutions we have $u_{2}^{\prime }\neq 0$
and from the first equation in (\ref{T22b}) we find $e^{2u_{2}}=\mathrm{const%
}\cdot \left\vert x+C\right\vert ^{\frac{4}{D}}$. With this function $%
u_{2}(x)$, the second equation gives $e^{2u_{0}}=\mathrm{const}\cdot
\left\vert x+C\right\vert ^{-\frac{2}{D}(D-2)}$. For these expressions of $%
u_{0}(x)$ and $u_{2}(x)$ the last equation in (\ref{T22b}) is obeyed
identically. Specifying the constants, the solution is presented in the Taub
form:%
\begin{equation}
ds_{\mathrm{T}}^{2}=\left\vert 1-\sigma x\right\vert ^{\frac{2}{D}\left(
2-D\right) }dt^{2}-dx^{2}-\left\vert 1-\sigma x\right\vert ^{\frac{4}{D}%
}\sum_{i=2}^{D}\left( dx^{i}\right) ^{2},  \label{s2Taub}
\end{equation}%
where $\sigma $ is another constant. This solution has a singularity at $%
x=1/\sigma $. For $D=3$ it is reduced to the Taub solution in General
Relativity. The higher dimensional generalization of the Taub solution has
also been considered in \cite{Gamb12}. For a test particle at rest with the
coordinate $x$, the acceleration in the geometry (\ref{s2Taub}) is expressed
as $a^{i}=\delta _{1}^{i}(1-2/D)/(x-1/\sigma )$. This corresponds to the
repulsion from the wall at $x=1/\sigma $ in both regions $x<1/\sigma $ and $%
x>1/\sigma $. Introducing the notations%
\begin{equation}
n_{D}=2\frac{D-1}{D},\;\sigma ^{\prime }=n_{D}^{\frac{D-2}{2(D-1)}}\sigma ,
\label{nD}
\end{equation}%
and new coordinates $x^{\prime i}$ in accordance with%
\begin{eqnarray}
t^{\prime } &=&n_{D}^{\frac{2-D}{2(D-1)}}t,\;1-\sigma ^{\prime }x^{\prime }=%
\frac{\left( 1-\sigma x\right) ^{n_{D}}}{n_{D}},  \notag \\
x^{\prime i} &=&n_{D}^{\frac{1}{D-1}}x^{i},\;i=2,\ldots ,D,  \label{tprime}
\end{eqnarray}%
the line element is written in the form%
\begin{equation}
ds_{\mathrm{T}}^{2}=\left\vert 1-\sigma ^{\prime }x^{\prime }\right\vert ^{%
\frac{2-D}{D-1}}\left( dt^{\prime 2}-dx^{\prime 2}\right) -\left\vert
1-\sigma ^{\prime }x^{\prime }\right\vert ^{\frac{2}{D-1}}\sum_{i=2}\left(
dx^{\prime i}\right) ^{2}.  \label{s2Taub3}
\end{equation}

As a simple example with two different metric tensors in the regions $x>0$
and $x<0$, we take $ds^{2}=ds_{\mathrm{T}}^{2}$ in the region $x<0$ (given
by (\ref{s2Taub}) with $\sigma >0$) and%
\begin{equation}
ds_{\mathrm{R}}^{2}=(1+x/b)^{2}dt^{2}-dx^{2}-\sum_{i=2}^{D}\left(
dx^{i}\right) ^{2},  \label{dsR2}
\end{equation}%
in the region $x>0$. The latter corresponds to the Rindler spacetime and is
obtained from (\ref{dsR}) redefining $x\rightarrow x+b$ and passing to a new
time coordinate $t=bt_{\mathrm{R}}$. For both regions $T_{\mathrm{(v)}%
i}^{k}=0$ and the metric tensor is regular. From (\ref{u02der}) one gets the
surface energy-momentum tensor required by the matching conditions: 
\begin{equation}
8\pi G\tau _{0}^{0}=-2\sigma \frac{D-1}{D},\;\tau _{1}^{1}=0,\;8\pi G\tau
_{2}^{2}=-\frac{1}{b}-\sigma \frac{D-2}{D}.  \label{tau}
\end{equation}%
Note that the corresponding energy density is negative. In the special case $%
\sigma =1/b$ we obtain $\tau _{2}^{2}=\tau _{0}^{0}$ and $\tau _{i}^{k}$
describes a CC-type source localized on the plane $x=0$.

\section{Solutions with cosmological constant}

\label{sec:CCsol}

In this section we consider the solutions of the gravitational field
equations (\ref{T22b}) with the CC $\Lambda $ as the only source. For the
corresponding energy-momentum tensor one has 
\begin{equation}
T_{i}^{k}=T_{(\Lambda )i}^{k}=\frac{\Lambda }{8\pi G}\delta _{i}^{k}.\;
\label{Tcc}
\end{equation}

\subsection{AdS spacetime}

For a negative CC from the first equation we have a special solution%
\begin{equation}
u_{2}^{\prime }=\pm \frac{1}{a},\;a=\sqrt{\frac{D\left( D-1\right) }{%
2|\Lambda |}}.  \label{u2dcc}
\end{equation}%
With this solution, the second equation in (\ref{T22b}) gives $u_{0}^{\prime
}=\pm 1/a$. The third equation is automatically satisfied. Fixing the
integration constants, the line element corresponding to this solution takes
the form%
\begin{equation}
ds^{2}=e^{\pm 2x/a}\left[ dt^{2}-\sum_{i=2}^{D}\left( dx^{i}\right) ^{2}%
\right] -dx^{2}.  \label{ds2AdS}
\end{equation}%
This line element describes AdS spacetime in Poincar\'{e} coordinates.
Introducing a new coordinate $z=\mp ae^{\mp x/a}$, $-\infty <\pm z<0$, the
line element is written in a conformally flat form%
\begin{equation}
ds^{2}=\frac{a^{2}}{z^{2}}\left[ dt^{2}-\sum_{i=2}^{D}\left( dx^{i}\right)
^{2}-dz^{2}\right] .  \label{ds2AdS2}
\end{equation}%
Here, the hypersurfaces $z=\mp \infty $ and $z=0$ correspond to the AdS
horizon and boundary, respectively. The acceleration of a test particle in
the geometry (\ref{ds2AdS}) is given by $a^{i}=\mp \delta _{1}^{i}/a$ and it
does not depend on the location of the particle. The latter property is a
consequence of the maximal symmetry of the AdS spacetime. The acceleration
is directed towards of the AdS horizon.

In the $D$-dimensional generalization of the Randall-Sundrum 1-brane model 
\cite{Rand99} the background line element reads%
\begin{equation}
ds^{2}=e^{-2|x|/a}\left[ dt^{2}-\sum_{i=2}^{D}\left( dx^{i}\right) ^{2}%
\right] -dx^{2},  \label{ds2RS}
\end{equation}%
and the brane is located at $x=0$. By taking into account that the volume
energy-momentum tensor is given by (\ref{Tcc}) in both regions $x<0$ and $%
x>0 $, from the matching conditions (\ref{u02der}) we get 
\begin{equation}
\tau _{0}^{0}=\tau _{2}^{2}=\frac{D-1}{4\pi Ga},\;\tau _{1}^{1}=0.
\label{tauRS}
\end{equation}%
This correspond to a positive CC localized on the brane.

\subsection{General solution for negative CC}

For a negative cosmological constant the first integral of the first
equation in (\ref{T22b}) is given by%
\begin{equation}
u_{2}^{\prime }=\frac{1}{a}\tanh w,\;w\equiv \frac{D(x-x_{0})}{2a},
\label{u2d}
\end{equation}%
with $x_{0}$ being an integration constant. Substituting this in the second
equation we get%
\begin{equation}
u_{0}^{\prime }=\frac{1}{2a}\left[ D\coth w-\left( D-2\right) \tanh w\right]
.  \label{u0d}
\end{equation}%
Now it can be checked that with these solutions for $u_{0}^{\prime }$ and $%
u_{2}^{\prime }$ the third equation in (\ref{T22b}) is obeyed identically.
The simple integration of (\ref{u2d}) and (\ref{u0d}) gives the functions $%
u_{0}(x)$ and $u_{2}(x)$. The corresponding line element reads%
\begin{equation}
ds^{2}=\frac{\sinh ^{2}w}{\left( \cosh w\right) ^{\frac{2}{D}(D-2)}}%
dt^{2}-dx^{2}-\left( \cosh w\right) ^{\frac{4}{D}}\sum_{i=2}^{D}\left(
dx^{i}\right) ^{2}.  \label{ds2ccn}
\end{equation}

Let us consider the asymptotic of the line element (\ref{ds2ccn}) for small
and large values of $|w|$. For $|w|\ll 1$, keeping the leading terms we get%
\begin{equation}
ds^{2}\approx x^{\prime 2}dt^{\prime 2}-dx^{\prime 2}-\sum_{i=2}^{D}\left(
dx^{i}\right) ^{2},  \label{ds2ccns}
\end{equation}%
where $x^{\prime }=x-x_{0}$ and $t^{\prime }=Dt/2a$. The right-hand side of (%
\ref{ds2ccns}) is the line element for the Rindler spacetime (compare with (%
\ref{dsR})). For large values of $|w|$, $|w|\gg 1$, keeping the leading
terms we get%
\begin{equation}
ds^{2}\approx e^{\pm 2x/a}\left[ dt^{\prime 2}-\sum_{i=2}^{D}\left(
dx^{\prime i}\right) ^{2}\right] -dx^{2},  \label{ds2ccnl}
\end{equation}%
with $t^{\prime }=2^{-2/D}e^{\mp x_{0}/a}t$ and $x^{\prime i}=2^{-2/D}e^{\mp
x_{0}/a}x^{i}$. Here, the upper and lower signs correspond to the cases $w>0$
and $w<0$, respectively. Hence, in this limit the asymptotic geometry
corresponds to the AdS spacetime.

For the acceleration of a test particle at rest one has $a^{i}=-\delta
_{1}^{i}u_{0}^{\prime }$ with $u_{0}^{\prime }$ given by (\ref{u0d}). It is
positive in the region $w<0$ and negative in the region $w>0$ and, hence,
the acceleration is directed towards the hyperplane $w=0$ which corresponds
to the Rindler horizon. At large distances from the horizon, corresponding
to $|w|\gg 1$, we get $a^{i}\approx -\delta _{1}^{i}\mathrm{sgn}(w)/a$. Near
the horizon the leading term in the asymptotic expansion is given by $%
a^{i}\approx \delta _{1}^{i}/(x_{0}-x)$. This term does not depend on the
value of CC.

\subsection{General solution for positive CC}

We turn to the case of $\Lambda >0$. By steps similar to those described in
the previous subsection we can show that%
\begin{equation}
u_{0}^{\prime }=\frac{1}{2a}\left[ D\cot w+\left( D-2\right) \tan w\right]
,\;u_{2}^{\prime }=-\frac{1}{a}\tan w.  \label{u0d2}
\end{equation}%
The further integrations of these relations lead to the line element%
\begin{equation}
ds^{2}=\frac{\sin ^{2}w}{\left\vert \cos w\right\vert ^{\frac{2}{D}(D-2)}}%
dt^{2}-dx^{2}-\left\vert \cos w\right\vert ^{\frac{4}{D}}\sum_{i=2}\left(
dx^{i}\right) ^{2}.  \label{ds2ccn2}
\end{equation}%
In this case the metric is a periodic function of $w$ with the period equal
to $\pi $. This corresponds to the periodicity with respect to the
coordinate $x$ with the period equal to $2\pi a/D$. The asymptotic of the
line element near the point $w=0$ is described by the Rindler line element (%
\ref{ds2ccns}). Near the point $w=\pi /2$ the line element is approximated
by the Taub solution:%
\begin{equation}
ds^{2}\approx \frac{dt^{2}}{\left\vert w-\pi /2\right\vert ^{\frac{2}{D}%
(D-2)}}-dx^{2}-\left\vert w-\pi /2\right\vert ^{\frac{4}{D}}\sum_{i=2}\left(
dx^{i}\right) ^{2}.  \label{ds2cc3}
\end{equation}%
Note that in this point we have a singularity.

By taking into account that the metric tensor is periodic, let us consider
the acceleration of the test particle, given as $a^{i}=-\delta
_{1}^{i}u_{0}^{\prime }$, in the region $-\pi /2<w<\pi /2$. It is positive
for $-\pi /2<w<0$ and negative for $0<w<\pi /2$. This means that, similar to
the case of negative CC, the acceleration is directed towards the Rindler
horizon $w=0$ with the near horizon asymptotic $a^{i}\approx \delta
_{1}^{i}/(x_{0}-x)$. The singular walls $w=\pm \pi /2$ are repulsive and
near of them the asymptoic of the acceleration is given by $a^{i}\approx
\delta _{1}^{i}(2-D)\mathrm{sgn}(w)/[a(\pi -2|w|)]$.

\section{CC slab with finite thickness}

\label{sec:CCslab}

As an application of the matching procedure and of the solutions described
above, here we consider a finite thickness slab with the CC energy-momentum
tensor (\ref{Tcc}) in the region $-L\leq x\leq L$. Different geometries in
the exterior regions $x<-L$ and $x>L$ will be discussed. Assuming a
symmetric configuration with respect to the plane $x=0$, firstly we consider
the interior line element (\ref{ds2ccn}) with $x_{0}=0$, corresponding to a
negative cosmological constant $\Lambda $. We have $ds^{2}=g_{ik}^{(\Lambda
)}dx^{i}dx^{k}$ with the metric tensor%
\begin{equation}
g_{ik}^{(\Lambda )}(x)=\mathrm{diag}\left( \frac{\sinh ^{2}w}{\left( \cosh
w\right) ^{\frac{2}{D}(D-2)}},-1,-\cosh ^{\frac{4}{D}}w,\cdots ,-\cosh ^{%
\frac{4}{D}}w\right) ,  \label{gikint}
\end{equation}%
in the region $|x|\leq L$ and $w=Dx/(2a)$. Rescaling the time and spatial
coordinates $x^{i}$, $i=2,\ldots ,D$, the interior line element can be
written in the form%
\begin{equation}
ds^{2}=g_{ik}(x)dx^{i}dx^{k},\;g_{ik}(x)=\frac{g_{ik}^{(\Lambda )}(x)}{%
|g_{ik}^{(\Lambda )}(L)|},\;|x|\leq L.  \label{gikint2}
\end{equation}%
With this normalization $g_{ik}(L)=\mathrm{diag}(1,-1,\ldots ,-1)$. This
normalization is convenient in the consideration of the matching conditions
discussed below.

\subsection{Minkowski exterior}

We start the discussion by the Minkowskian geometry in the exterior regions:%
\begin{equation}
ds_{\mathrm{M}}^{2}=dt^{2}-dx^{2}-\sum_{i=2}^{D}\left( dx^{i}\right)
^{2},\;|x|>L.  \label{dsM}
\end{equation}%
With the choice (\ref{gikint2}) for the interior line element, the metric
tensor is continuous on the boundaries $x=\pm L$. The surface
energy-momentum tensor is determined by the matching conditions (\ref{u02der}%
). By taking into account that 
\begin{equation}
u_{0}^{\prime }=\frac{1}{2a}\left[ D\coth w-\left( D-2\right) \tanh w\right]
,\;u_{2}^{\prime }=\frac{1}{a}\tanh w,  \label{u02deri}
\end{equation}%
in the region $|x|<L$ and $u_{0}^{\prime }=u_{2}^{\prime }=0$ for $|x|>L$,
from (\ref{u02der}) we get%
\begin{equation}
\tau _{0}^{0}=\frac{D-1}{8\pi Ga}\tanh w_{L},\;\tau _{1}^{1}=0,\;\tau
_{2}^{2}=\frac{D-2}{D-1}\frac{\tau _{0}^{0}}{2}+\frac{D\coth w_{L}}{16\pi Ga}%
,  \label{tauM}
\end{equation}%
where $w_{L}=DL/(2a)$. The surface energy density and the stresses (no
summation over $i$) $\tau _{i}^{i}=\tau _{2}^{2}$, $i=2,\ldots ,D$, are
positive. Note that the effective pressure along the $i$th spatial direction
is given by $-\tau _{i}^{i}$ and in the example under consideration it is
negative.

\subsection{Rindler exterior}

For the exterior Rindler geometry the line element is given by (\ref{dsR})
in the regions $|x|>L$ and for the interior geometry we have (\ref{gikint2}%
). Introducing a new Rindler time coordinate $t$ in accordance with $t=Lt_{%
\mathrm{R}}$, we see that the metric tensor is continuous on the boundaries $%
x=\pm L$. The derivatives in the matching conditions are given by (\ref%
{u02deri}) in the region $|x|<L$ and by $u_{0}^{\prime }=1/x$, $%
u_{2}^{\prime }=0$ in the region $|x|>L$. From (\ref{u02der}) one finds%
\begin{eqnarray}
\tau _{0}^{0} &=&\frac{D-1}{8\pi Ga}\tanh w_{L},\;\tau _{1}^{1}=0,  \notag \\
\tau _{2}^{2} &=&\frac{D-2}{D-1}\frac{\tau _{0}^{0}}{2}+D\frac{\coth
w_{L}-1/w_{L}}{16\pi Ga}.  \label{tauR}
\end{eqnarray}%
The surface energy density is the same as that for the Minkowski exterior,
whereas the stresses are different. Note that, depending on the value of the
parameter $w_{L}$, the effective pressure $-\tau _{2}^{2}$ can be either
negative or positive.

\subsection{Taub exterior}

The exterior geometry is described by the line element (\ref{s2Taub}).
Redefining the coordinates and the constant $\sigma $, we rewrite it in the
form%
\begin{equation}
ds_{\mathrm{T}}^{2}=\left( \frac{1+\sigma |x|}{1+\sigma L}\right) ^{\frac{2}{%
D}(2-D)}dt^{2}-dx^{2}-\left( \frac{1+\sigma |x|}{1+\sigma L}\right) ^{\frac{4%
}{D}}\sum_{i=2}^{D}\left( dx^{i}\right) ^{2}.  \label{s2Taub2}
\end{equation}%
For $\sigma >0$ the metric tensor is regular. With this line element in the
region $|x|>L$ and with the line element (\ref{gikint2}) for $|x|<L$, the
metric tensor is continuous at $x=\pm L$. By taking into account that%
\begin{equation}
u_{0}^{\prime }=\frac{2-D}{D}\frac{\sigma \,\mathrm{sgn}(x)}{1+\sigma |x|}%
,\;u_{2}^{\prime }=\frac{2}{D}\frac{\sigma \,\mathrm{sgn}(x)}{1+\sigma |x|},
\label{u02dT}
\end{equation}%
for $|x|>L$, the matching conditions at $x=L$ give%
\begin{eqnarray}
\tau _{0}^{0} &=&\frac{D-1}{8\pi Ga}\left( \tanh w_{L}-\frac{2}{D}\frac{%
\sigma a}{1+\sigma L}\right) ,  \notag \\
\tau _{2}^{2} &=&\frac{D-2}{D-1}\frac{\tau _{0}^{0}}{2}+\frac{D\coth w_{L}}{%
16\pi Ga},  \label{tauT}
\end{eqnarray}%
and $\tau _{1}^{1}=0$. Note that for the Taub exterior, depending on the
relative values of $a$ and $L$, the surface energy density can be either
positive or negative.

\subsection{Slab with positive CC}

Now we turn to the slab with positive CC. The interior line element is given
by (\ref{ds2ccn2}) with the metric tensor%
\begin{equation}
g_{ik}^{(\Lambda )}(x)=\mathrm{diag}\left( \frac{\sin ^{2}w}{|\cos w|^{2%
\frac{D-2}{D}}},-1,-|\cos w|^{\frac{4}{D}},\cdots ,-|\cos w|^{\frac{4}{D}%
}\right) ,  \label{gikintp}
\end{equation}%
where $w=Dx/(2a)$. Again, rescaling the coordinates the line element is
presented in the form (\ref{gikint2}) with $g_{ik}^{(\Lambda )}(L)=\mathrm{%
diag}(1,-1,\ldots ,-1)$. We will assume that $L<\pi a/D$. In this case the
metric tensor is regular inside the slab. The derivatives of the functions $%
u_{0}(x)$ and $u_{2}(x)$ in the region $|x|<L$ are given by (\ref{u0d2}).
For the case of $\Lambda >0$, the components of the surface energy-momentum
tensor for the exterior Minkowski, Rindler and Taub geometries are obtained
from the formulas given above for $\Lambda <0$ by the replacements%
\begin{equation}
\tanh w_{L}\rightarrow -\tan w_{L},\;\coth w_{L}\rightarrow \cot w_{L}.
\label{Repl}
\end{equation}%
The surface energy density is negative for all those geometries.

\section{Conclusion}

\label{sec:Conc}

We have considered plane symmetric solutions of General Relativity for
general number of spatial dimensions. For the metric tensor given by (\ref%
{ds2s}), the field equations are presented in the form (\ref{T22}) and the
covariant continuity equation for the energy-momentum tensor is reduced to (%
\ref{Cons}). The set of gravitational equations is simplified by the choice
of the coordinate $x$ in accordance with (\ref{u10}). By using those
equations one can derive the matching conditions for the metric tensor in
the problems where the geometry is described by two distinct line elements
in neighboring half-spaces. The metric tensor is continuous on the
separating boundary and the discontinuity of its first order derivative is
given by (\ref{u02der}), where $\tau _{i}^{k}$ is the surface
energy-momentum tensor.

Two classes of the vacuum solutions of the gravitational field equation are
presented. The first one corresponds to the Rindler spacetime and the second
one is a higher-dimensional generalization of the well known Taub solution.
By an appropriate choice of the integration constants the latter is given by
(\ref{s2Taub}) (see also \cite{Gamb12}). It has a singularity at $x=1/\sigma 
$ that presents a repulsive wall for test particles. As a simple example of
geometry with two distinct metric tensors in two different regions we have
considered the combination of the Rindler and Taub geometries separated by a
planar boundary. The components of the corresponding surface energy-momentum
tensor are expressed by (\ref{tau}).

As an example of the source in gravitational field equations we have
considered the CC $\Lambda $. For negative CC there is a special solution
that corresponds to $(D+1)$-dimensional AdS spacetime. In Poincar\'{e}
coordinates the line element has the form (\ref{ds2AdS}). In the
Randall-Sundrum 1-brane model two copies of the AdS half-space are combined
in the form of Eq. (\ref{ds2RS}). The surface energy-momentum tensor on the
separating brane is given by (\ref{tauRS}). The general solutions of the
field equations for negative and positive CC are given by (\ref{ds2ccn}) and
(\ref{ds2ccn2}), respectively. In the case of a negative CC the geometry is
non-singular. For small and large values of the variable $|w|$ it is
approximated by the Rindler and AdS spacetimes, respectively. For a positive
CC the metric tensor is a periodic function of $x$ with the period $2\pi a/D$%
. In this case one has singularities at the points corresponding to $%
w=(n+1/2)\pi $. Near these points the geometry is approximated by the Taub
solution. For both solutions with negative and positive CC the hyperplane $%
w=0$ ($x=x_{0}$) corresponds to a horizon that is the analog of the Rindler
horizon. The acceleration of a test particle at rest is directed towards the
horizon.

By using the solutions with a CC we have constructed a simple model of a
finite thickness slab symmetric with respect to the central plane. The
volume energy-momentum tensor inside the slab is given by (\ref{Tcc}) and in
the exterior regions we have used the vacuum solutions of the field
equations. Three different cases have been considered with the Minkowski,
Rindler and Taub geometries. For the latter geometry the singularity-free
Taub solution is employed. The corresponding surface energy-momentum tensors
are expressed by (\ref{tauM}), (\ref{tauR}) and (\ref{tauT}), respectively.
For a slab with positive CC the interior geometry is non-singular for the
half-thickness obeying the condition $L<\pi a/D$.

The setup considered in the present paper can be used for the investigation
of the backreaction effects of the vacuum polarization of quantum fields
induced by boundaries with $x=\mathrm{const}$. The boundary conditions
imposed on quantum fields lead to the modification of the spectrum for
vacuum fluctuations and, as a consequence, the vacuum expectation values of
physical observables are changed. In particular, the vacuum energy-momentum
tensor for planar boundaries has been widely considered in the literature.
The simplest example is the Casimir effect (see, for example, \cite{Bord09})
for perfectly conducting parallel plates in the Minkowski spacetime. Already
in that simple example the vacuum stresses are anisotropic. The planar
boundaries in the Rindler spacetime, corresponding to uniformly accelerated
plates in the Fulling-Rinlder vacuum, have been considered in \cite{Cand77}-%
\cite{Saha04}. The references for the corresponding investigations in the
AdS bulk can be found in \cite{Saha20}.

\section*{Acknowledgments}

The work was supported by the grant No. 21AG-1C047 of the Higher Education
and Science Committee of the Ministry of Education, Science, Culture and
Sport RA.

\end{document}